\documentclass{emulateapj}
\shorttitle{A globular cluster in M31}
\shortauthors{Ma et al.}

\begin{document}
\slugcomment{ApJ, in press}
\title{Age constraints for an  M31 globular cluster from SEDs-fit}

\author{
Jun Ma\altaffilmark{1},  Yanbin Yang\altaffilmark{1}, David
Burstein\altaffilmark{2}, Zhou Fan\altaffilmark{1,3}, Zhenyu
Wu\altaffilmark{1}, Xu Zhou\altaffilmark{1}, Jianghua
Wu\altaffilmark{1}, Zhaoji Jiang\altaffilmark{1} and Jiansheng
Chen\altaffilmark{1}}

\altaffiltext{1}{National Astronomical Observatories, Chinese
Academy of Sciences, Beijing, 100012, P. R. China;
majun@vega.bac.pku.edu.cn}

\altaffiltext{2}{Department of Physics and Astronomy, Box 871504,
Arizona State University, Tempe, AZ 85287--1504}

\altaffiltext{3}{Graduate University of Chinese Academy of
Sciences}

\begin{abstract}

We have constrained the age of the globular cluster S312 in the
Andromeda galaxy (M31) by comparing its multicolor photometry with
theoretical stellar population synthesis models. This is both a
check on the age of this globular cluster, as well a check on our
methodology. Main-sequence photometry has been the most direct
method for determining the age of a star cluster. S312 was
observed as part of the Beijing-Arizona-Taiwan-Connecticut (BATC)
Multicolor Sky Survey from 1995 February to 2003 December. The
photometry of BATC images for S312 was taken with 9
intermediate-band filters covering $5000-10000$\AA. Combined with
photometry in the near-ultraviolet (NUV) of {\sl GALEX},
broad-band $UBVR$ and infrared $JHK_s$ of 2MASS, we obtained the
accurate spectral energy distributions (SEDs) of S312 from
$2267-20000$\AA. A quantitative comparison to simple stellar
population models yields an age of $9.5_{-0.99}^{+1.15}$~Gyr,
which is in very good agreement with the previous determination by
main-sequence photometry. S312 has a mass of $9.8\pm{1.85}\times
10^5 \rm M_\odot$, and is a medium-mass globular cluster in M31.
By analysis of errors of ages determined based on the SED fitting
method of this paper, secure age constraints are derived with
errors of $< 3$ Gyr for ages younger than 9 Gyr. In fact, the
theoretical SEDs are not sensitive to the variation of age for
ages greater than $\sim 10$ Gyr. Therefore, for globular clusters
as old as the majority of the Galactic GCs, our method do not
distinguish them accurately. We emphasize that our results show
that even with multiband photometry spanning NUV to $K_s$, our age
constraints from SED fitting are distressingly uncertain, which
has implications for age derivations in extragalactic globular
cluster systems.

\end{abstract}


\section{Introduction}

Galactic globular clusters (GCs), which are thought to be among
the oldest stellar objects in the Universe, provide vitally
important information regarding the minimum age of the Universe
and the early formation history of our Galaxy. The most direct
method for determining the age of a star cluster is main-sequence
photometry, since the turn-off is mostly affected by age
\citep[see][and references therein]{puzia02b}. However, this
method has only been applied to the Galactic GCs and globular
clusters in the satellites of the Milky Way
\citep[e.g.,][]{rich01} before \citet{brown04}. Generally,
extragalactic globular cluster ages are inferred from composite
colors and/or spectroscopy. S312, first detected by \citet{sh73}
(No.19), then by \citet{sarg77} (No.312=S312) and \citet{battis87}
(No.379=Bo379), is among the first extragalactic globular clusters
whose age was accurately estimated by main-sequence photometry
\citep{brown04}. \citet{brown04} obtained the color-magnitude
diagram (CMD) below the main-sequence turnoff for S312 using the
extremely deep images of M31 with the Advanced Camera for Surveys
(ACS) on the $Hubble$ $Space$ $Telescope$ (HST). \citet{brown04}
estimated an age of 10 Gyr for S312 by the quantitative comparison
to isochrones of \citet{vandenberg06}.

This cluster was also observed as part of the galaxy calibration
program of the Beijing-Arizona-Taiwan-Connecticut (BATC)
Multicolor Sky Survey \citep[e.g.,][]{fan96,zheng99} in 9
intermediate-band filters. Combined with photometry in NUV of {\sl
GALEX}, broad-band $UBVR$ and infrared $JHK_s$ of the Two Micron
All Sky Survey (2MASS), we obtained the accurate SEDs of S312 in
17 filter bands from $2267-20000$\AA.

Since the pioneering work of \citet{Tinsley68,Tinsley72} and
\citet{SSB73}, evolutionary population synthesis modelling has
become a powerful tool to interpret integrated spectrophotometric
observations of galaxies and their subcomponents, such as star
clusters \citep{Anders04}. The evolution of star clusters is
usually modelled by means of a simple stellar population (SSP)
approximation. An SSP is defined as a single generation of coeval
stars formed from the same progenitor molecular cloud (thus
implying a single metallicity), and governed by a given initial
mass function (IMF). Globular clusters, which are bright and
easily identifiable, are typically characterized by homogeneous
abundance and age distributions. For example, \citet{bh01}
compared the predicted SSP colors of three stellar population
synthesis models to the intrinsic broad-band $UBVIRJHK$ colors of
Galactic and M31 GCs, and found that the best-fitting models match
the clusters' SEDs very well indeed.

In this paper, we constrain the age of S312 by comparing
observational SEDs (Sect. 2) with population synthesis models in
Sect. 3. Our independently-constrained result is in very good
agreement with the previous determination of \citet{brown04}. We
give discussions and summarize our results in Sect. 4.

\section{Optical and infrared observations of GC S312}

\subsection{Historical observations of GC S312}

S312 was first detected by \citet{sh73}, who searched 25 M31
globular clusters and cluster candidates using plates taken with
the 80/120/240-cm Schmidt telescope of the Radio Astrophysical
Observatory of the Latvian Academy of Sciences. Since that time,
\citet{sarg77} and \citet{battis87} confirmed S312 to be an M31
globular cluster.

\subsection{Archival images of the BATC sky survey}

Observations of S312 were obtained by the BATC 60/90cm Schmidt
telescope located at the XingLong station of the National
Astronomical Observatory of China (NAOC). This telescope has 15
intermediate-band filters covering the optical wavelength range of
$3000-10000$ \AA, and is specifically designed to avoid
contaminations from the brightest and most variable night-sky
emission lines. Descriptions of the BATC photometric system can be
found in \citet{fan96}. The finding chart of GC S312 in the $g$
band (centered on 5795 \AA) of the BATC system with the NAOC
60/90cm Schmidt telescope is illustrated in Figure 1.

\begin{figure*}
\figurenum{1} \hspace{1.0cm}\rotatebox{0}{\plotone{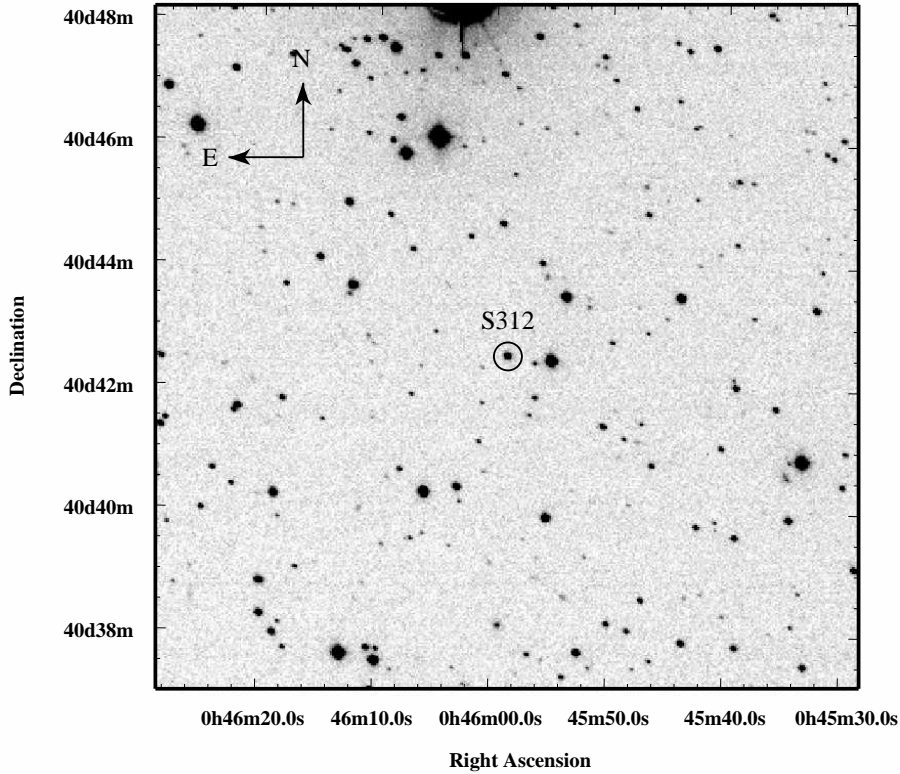}}
\vspace{0cm} \caption{An image of S312 in the BATC $g$ band of the
NAOC 60/90cm Schmidt telescope. S312 is circled. The field of view
of the image is $11^{\prime}\times 11^{\prime}$.} \label{fig:one}
\end{figure*}

We extracted 125 images of M31, taken in 9 BATC filters from the
BATC survey archive during 1995 February--2003 December. Table 1
contains the log of observations. Multiple images of the same
filter were combined to improve the image quality.

\subsection{Intermediate-band photometry of S312}

Since S312 is located in the M31 halo, it is easy to do the
photometry. The intermediate-band magnitudes of S312 are determined on
the combined images with the standard aperture photometry, i.e. the
PHOT routine in DAOPHOT \citep{stet87}. The BATC photometric system
calibrates the magnitude zero level similar to the spectrophotometric
AB magnitude system. For the flux calibration, the Oke-Gunn primary
flux standard stars HD 19445, HD 84937, BD +26$^{\circ}$2606, and BD
+17$^{\circ}$4708 taken from \citet{ok83}, were observed during
photometric nights \citep{yan00}. The results of well-calibrated
photometry of S312 in 9 filters are summarized in the fifth column of
Table 1.

\subsection{{\sl GALEX}, broad-band and 2MASS photometry of S312}

In order to constrain the age of S312 accurately, we use as many
photometric data points covering as large a wavelength range as
possible. M31 field was observed as part of the Nearly Galaxy
Survey (NGS) carried out by {\sl Galaxy Evolution Explorer} ({\sl
GALEX}) in two UV bands: far-ultraviolet (FUV) and
near-ultraviolet (NUV) \citep[see details from][]{rey05,rey06}.
\citet{rey06} presented the photometric data for 485 and 273 M31
GCs in {\sl GALEX} NUV and FUV, respectively. S312 was not
detected in FUV from \citet[][also S.-C. Rey, priv. comm.]{rey06}.
Using the CCD imaging of 0.9m telescope at the Kitt Peak National
Observatory, \citet{rhh94} presented CCD $BVR$ integrated
magnitudes and color indices for 41 globular clusters and cluster
candidates in the outer halo of M31 including S312.
\citet{battis87} presented $UBVr$ photometry from the photographic
plates for their most sample globular clusters and candidates
including S312, however they did not give photometric
uncertainties. In this paper, for S312, we adopted CCD $BVR$
photometry of \citet{rhh94}, and photographic $U$ photometry of
\citet{battis87} with a photometric uncertainty of 0.08 mag as
suggested by \citet{gall04}.

As pointed out by \citet{wor94}, the age-metallicity degeneracy in
optical broad-band colors is $\rm{\Delta age/\Delta Z\sim 3/2}$,
implying that the composite spectrum of an old stellar population
is indistinguishable from that of a younger but more metal-rich
population (and vice versa) \citep[also see][]{MacArthur04}.
\citet{jong96} showed that this degeneracy can be partially broken
by adding 2MASS infrared photometry to the optical colors.
\citet{Cardiel03} found that inclusion of an infrared band can
improve the predictive power of the stellar population diagnostics
by $\sim 30$ times over using optical photometry alone.
\citet{wu05} also showed that the use of near-infrared colors can
break the age-metallicity degeneracy. At the same time, as
\citet{kbm02} and \citet{puzia02a} indicated that near-infrared
photometry is less sensitive to interstellar extinction with
respect to the classical optical bands, it provides useful
complementary information that can help to disentangle the
age-metallicity degeneracy \citep[also see][]{gall04}.

In this paper, we add $JHK_s$ photometry from the 2MASS Point
Source Catalogue (PSC) to broad- and intermediate-band optical,
and ultraviolet photometry to constrain the age of S312 to the
highest possible accuracy. 2MASS presents $J,H$ and $K_s$ complete
homogeneous photometry of 99.998 \% of the sky down to $K_s\sim
15.5$ (completeness $>$ 99 \% for $K_s\leq 14.3$). The
near-ultraviolet, broad-band and 2MASS photometric data points of
S312 are listed in Table 2.

\section{Stellar population of S312}

\subsection{Stellar populations and synthetic photometry}

As a check on the age of S312, and also on a check of our
methodology, we compare its spectral energy distributions with
theoretical stellar population synthesis models. We used the SSP
models of \citet{bru03} (hereafter BC03), which have been upgraded
from the \citet{bc93,bc96} version, and now provide the evolution
of the spectra and photometric properties for a wider range of
stellar metallicities. BC03 provide 26 SSP models (both of high
and low resolution) using the 1994 Padova evolutionary tracks, 13
of which were computed using the \citet{chabrier03} IMF assuming
lower and upper mass cut-offs of $m_{\rm L}=0.1~M_{\odot}$ and
$m_{\rm U}=100~M_{\odot}$, respectively. The other 13 were
computed using the \citet{salp55} IMF with the same mass cut-offs.
In addition, BC03 provide 26 SSP models using the 2000 Padova
evolutionary tracks. However, as \citet{bru03} pointed out, the
2000 Padova models, which include more recent input physics than
the 1994 models, tend to produce worse agreement with observed
galaxy colors. These SSP models contain 221 spectra describing the
spectral evolution of SSPs from $1.0\times10^5$ yr to 20 Gyr. The
evolving spectra include the contribution of the stellar component
at wavelengths from 91\AA~~to $160\mu$m. In this paper, we adopt
the high-resolution SSP models computed using the 1994 Padova
evolutionary tracks and a \citet{salp55} IMF\footnote{We note that
because of the slow SED evolution of SSPs at ages in excess of a
few Gyr, all of the most commonly used spectral synthesis models
agree very well at these ages. Therefore, the choice of IMF is
{\it only} important for estimating the photometric mass of the
cluster, and does {\it not} affect the determination of the age of
S312.}. We note that although the current best constraint of
the age of the Universe is of order 13.7 Gyr, the SSP models and
the stellar evolutionary tracks that form their basis have been
calculated for ages up to 20 Gyr. It is not straightforward to
correct for this discrepancy; one would need to recalculate all
stellar evolutionary tracks for all metallicities.

Since our observational data are integrated luminosites through
our set of filters, we convolved the BC03 SSP SEDs with the BATC
intermediate-, {\sl GALEX} NUV-, broad-band $UBVR$ and 2MASS
filter response curves to obtain synthetic ultraviolet, optical
and near-infrared photometry for comparison. The synthetic $i{\rm
th}$ filter magnitude can be computed as

\begin{equation}
m=-2.5\log\frac{\int_{\lambda}F_{\lambda}\varphi_{i} (\lambda){\rm
d}\lambda}{\int_{\lambda}\varphi_{i}(\lambda){\rm
d}\lambda}-48.60,
%\quad,
\end{equation}
where $F_{\lambda}$ is the theoretical SED and $\varphi_{i}$ the
response curve of the $i{\rm th}$ filter of the BATC, {\sl GALEX},
$UBVR$ and 2MASS photometric systems. Here, $F_{\lambda}$ varies
with age and metallicity.

\subsection{Reddening and metallicity of S312}

To obtain the intrinsic SEDs of S312, the photometry must be
dereddened. \citet{bh00} determined the reddening for each
individual cluster using correlations between optical and infrared
colours and metallicity and by defining various ``reddening-free''
parameters using their large database of multi-colour photometry.
\citet{bh00} found that the M31 and Galactic GC extinction laws,
and the M31 and Galactic GC colour-metallicity relations are
similar to each other. They then estimated the reddening to M31
objects with spectroscopic data using the relation between
intrinsic optical colours and metallicity for Galactic clusters.
For objects without spectroscopic data, they used the
relationships between the reddening-free parameters and certain
intrinsic colours based on the Galactic GC data. \citet{bh00}
compared their results with those in the literature and confirmed
that their estimated reddening values are reasonable, and
quantitatively consistent with previous determinations for GCs
across the entire M31 disc. In particular, \citet{bh00} showed
that the distribution of reddening values as a function of
position appears reasonable in that the objects with the smallest
reddening are spread across the disk and halo, while the objects
with the largest reddening are concentrated in the galactic disk.
Therefore, we adopted the reddening values from \citet[][also P.
Barmby, priv. comm.]{bh00} for S312 to be $E(B-V)=0.10$ as
\citet{rey06} did. The values of extinction coefficient
$R_{\lambda}$ are obtained by interpolating the interstellar
extinction curve of \citet{car89}.

The SEDs for clusters are significantly affected by the adopted
metallicity, especially for old clusters. So, only if the
metallicity is known, the age of a cluster can be constrained
accurately. \citet{hbk91} derived metallicities for 150 M31
globular clusters including S312, using the strengths of six
absorption features in the cluster integrated spectra. The
metallicity of S312, obtained by \citet{hbk91}, is $\rm
[Fe/H]=-0.7$. \citet{hfr97} used $HST$ WFPC2 photometry to
construct deep color-magnitude diagrams for S312, and the shape of
the red giant branch gave an iron abundance of $\rm [Fe/H]=-0.53$.
In this paper, we adopt for comparison $\rm [Fe/H]=-0.53$ as
\citet{brown04} did.

\subsection{Comparison of the age scale}

Before constraining the age of S312, we compare the age
scale between the Padova 1994 evolutionary tracks
\citep{bertelli94} used by BC03 and the 2006 vandenBerg isochrones
\citep{vandenberg06} used in \citet{brown04}. As an example, we
only draw the isochrones with 10 Gyr and the solar metallicity.
Figure 2 shows the isochrones between effective temperaure and
bolometric magnitudes. It is clear that the matching is very good
in the main sequence (MS) and the subgiant branch (SGB). As we
know, the main-sequence photometry focuses on the upper end of the
MS and SGB, since uncertainties in models of stars on the red
giant branch (RGB) are large. So, if \citet{brown04} had fitted
the color-magnitude diagram (CDM) of S312 with the Padova 1994
evolutionary tracks, its age would have been close to 10 Gyr.

\begin{figure*}
\figurenum{2} \epsscale{1.0} \hspace{0cm}{\plotone{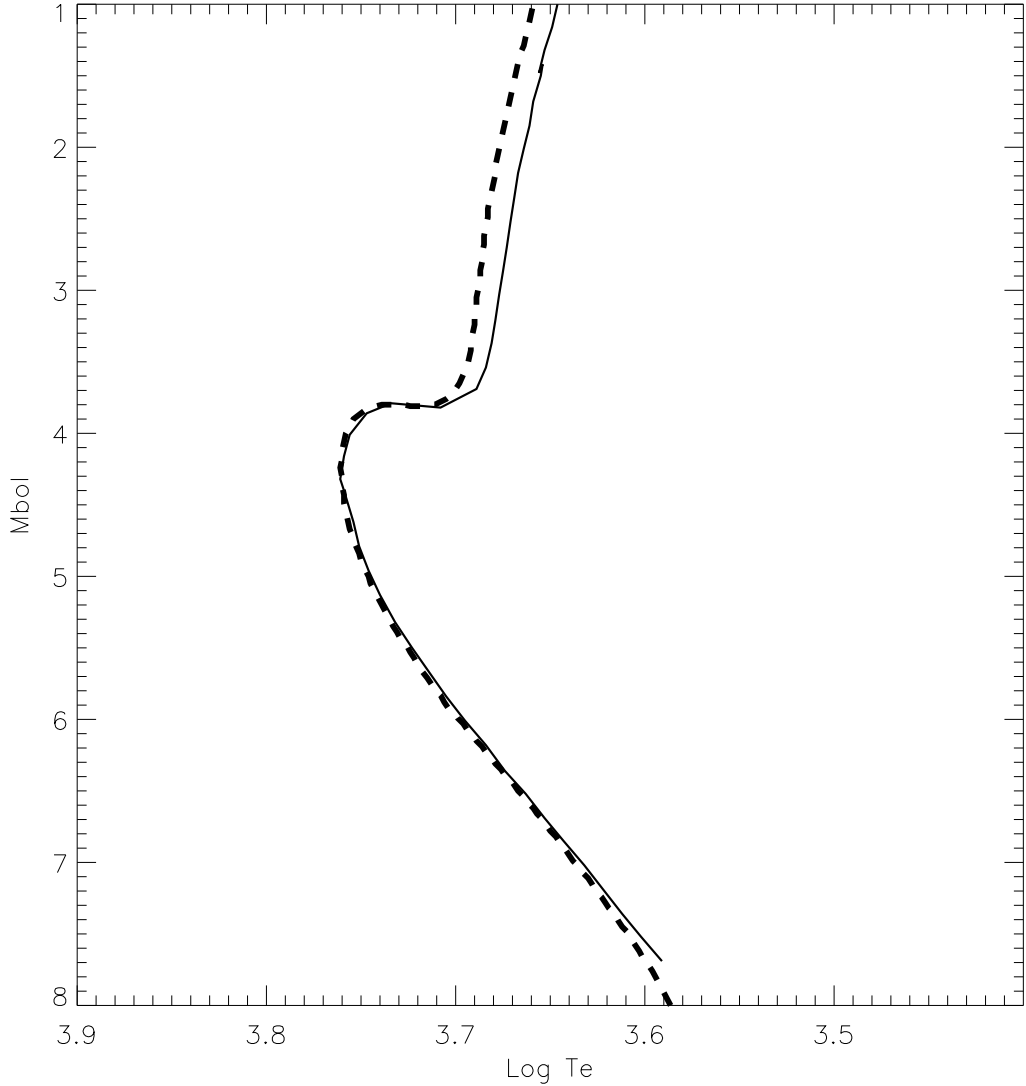}}
\vspace{0.0cm} \caption{Comparison of the age scale between the
Padova 1994 tracks and the 2006 vandenBerg isochrones used in
\citet{brown04}. Dashed and solid lines are the Padova 1994 tracks
and 2006 vandenBerg isochrone, respectively, for age 10 Gyr and
the solar metallicity.} \label{fig:two}
\end{figure*}

\subsection{Fit results}

We use a $\chi^2$ minimization test to examine which BC03 SSP
models are most compatible with the observed SEDs, following

\begin{equation}
\chi^2=\sum_{i=1}^{17}{\frac{[m_{\lambda_i}^{\rm
intr}-m_{\lambda_i}^{\rm mod}(t)]^2}{\sigma_{i}^{2}}},
%\quad,
\end{equation}
where $m_{\lambda_i}^{\rm mod}(t)$ is the integrated magnitude in
the $i{\rm th}$ filter of a theoretical SSP at age $t$,
$m_{\lambda_i}^{\rm intr}$ presents the intrinsic integrated
magnitude in the same filter and

\begin{equation}
\sigma_i^{2}=\sigma_{{\rm obs},i}^{2}+\sigma_{{\rm mod},i}^{2}.
%\quad .
\end{equation}
Here, $\sigma_{{\rm obs},i}^{2}$ is the observational uncertainty,
and $\sigma_{{\rm mod},i}^{2}$ is the uncertainty associated with
the model itself, for the $i{\rm th}$ filter. \citet{charlot96}
estimated the uncertainty associated with the term $\sigma_{{\rm
mod},i}^{2}$ by comparing the colors obtained from different
stellar evolutionary tracks and spectral libraries. Following
\citet{wu05}, we adopt $\sigma_{{\rm mod},i}^{2}=0.05$ in this
paper.

The BC03 SSP models include six initial metallicities, ${\rm
[Fe/H]}=-2.2490$, $-1.6464$, $-0.6392$, $-0.3300$, $+0.0932$
(solar metallicity), and $+0.5595$. Spectra for other
metallicities can be obtained by linear interpolation of the
appropriate spectra for any of these metallicities. For S312,
whose metallicity and reddening values were published by other
authors, the cluster age is the sole parameter to be estimated
(for a given IMF and extinction law, which we assume to be
universal). In Figure 3, we show the intrinsic SEDs of S312, the
integrated SEDs of the best-fitting model, and the spectra of the
best-fitting model superimposed the 3-sigma spectra. The best
reduced $\chi^2$ of 18.23 is achieved with an age of
$9.5_{-0.99}^{+1.15}$~Gyr, which is in very good agreement with
the previous determination (10 Gyr) of \citet{brown04} by
main-sequence photometry. From Figure 3, we can see that, the
3-sigma spectra deviate clearly from best-fitting ones, i.e. these
three spectra clearly separate. This indicates that we can
constrain the age of S312 with errors of about 3 Gyr.

\begin{figure*}
\figurenum{3} \epsscale{1.0} \hspace{0cm} {\plotone{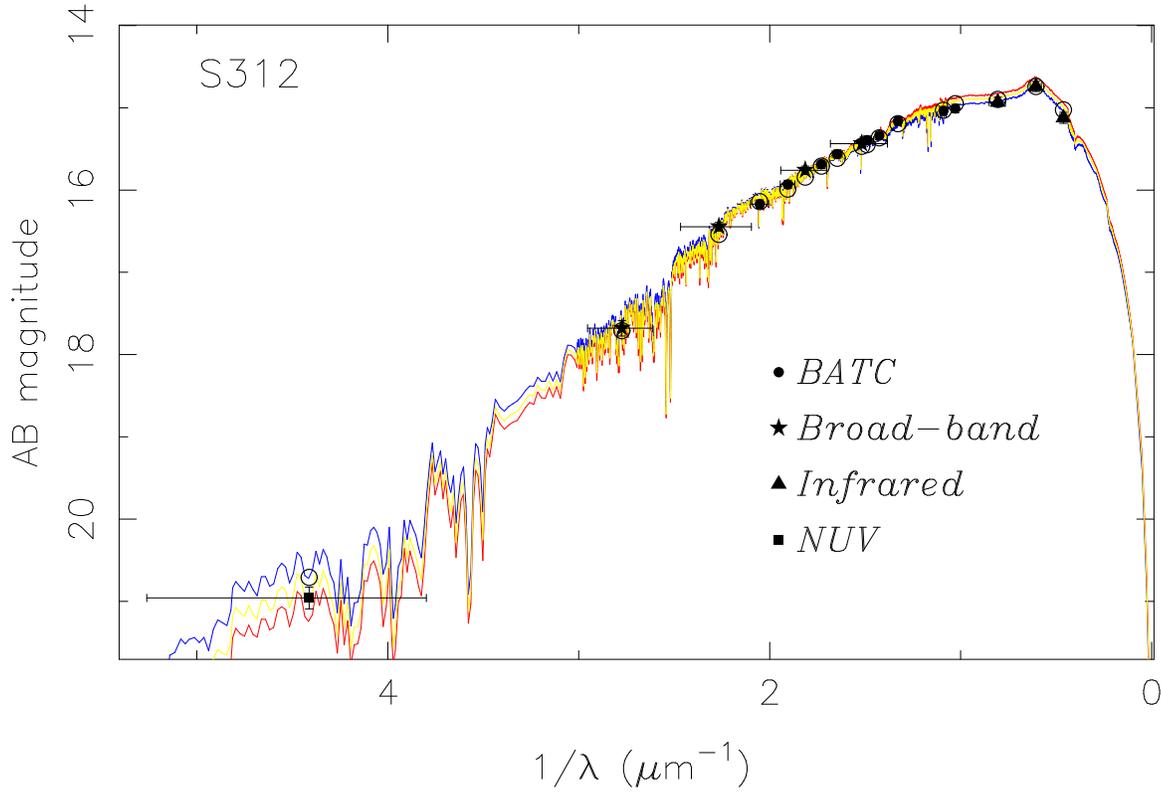}}
\vspace{0.0cm} \caption{Plots of the best-fitting integrated SEDs
of SSP models plotted on top of the intrinsic SEDs for S312. The
photometry itself is shown by the symbols with error bars
(vertical ones for uncertainties and horizontal ones for the
approximate wavelength coverage of each filter). Open circles
represent the calculated magnitude of the model SED for each
filter. Yellow, red and blue indicate the spectra of the
best-fitting model, plus and minus the $3\sigma$ spectra,
respectively.} \label{fig:three}
\end{figure*}

\subsection{Mass of S312}

Star cluster masses can be estimated by comparing the measured
luminosity in $V$ band with the theoretical mass-to-light ratios.
These ratios are a function of the cluster age and metallicity.
BC03 calculated these ratios for a range of metallicities.
Mass-to-light ratios for other metallicities can be obtained by
linear interpolation of the appropriate ratios for any of these
metallicities. For $\rm [Fe/H]=-0.53$ and 9.5~Gyr, we derived the
mass-to-light ratio in broad-band $V$ to be $3.69~\rm
M_\odot/L_\odot$. Based on its present luminosity, $V = 16.13 \pm
0.01$ mag and the extinction, $E(B-V) = 0.10$, its intrinsic
luminosity is $V_0 = 15.82 \pm 0.01$ mag [assuming the
\citet{car89} Galactic reddening law; $A_V = 0.31$ mag]. From
this, we determine the mass of S312 to be $9.8\pm{1.85}\times 10^5
\rm M_\odot$. Comparing with 037-B327 [$\mathcal{M}_{\rm 037-B327}
\sim 8.5 \times 10^6$ M$_\odot$ \citep{bk02} or $\mathcal{M}_{\rm
037-B327} \sim 3.0 \pm 0.5 \times 10^7$ M$_\odot$ \citep{Ma06}]
and G1 [$\mathcal{M}_{\rm G1} \sim (7-17)\times 10^6$ M$_\odot$
\citep{meylan01}] in M31 and $\omega$ Cen
[$\mathcal{M}_{\omega{\rm~Cen}} \sim (2.9 - 5.1) \times
10^6$M$_\odot$ \citep{meylan02}] in the Milky Way, the most
massive clusters in the Local Group, S312 is only a medium-mass
globular cluster.

\subsection{Comparison with SSP models based on the 2000 Padova
evolutionary tracks}

As discussed in Section 3.1, BC03 provide two sets of SSP models
based on the 1994 and 2000 Padova evolutionary tracks,
respectively. In Section 3.4, we adopted the SSP models based on
the 1994 Padova evolutionary tracks to constrain the age of S312.
However, it is interesting to show what the best fit is based on
the 2000 Padova models. Using the 2000 Padova models, we fitted
the intrinsic SEDs of S312 again, it is surprised that the fitting
is very good with the best reduced $\chi^2$ of 9.71, but, the age
of $11.75_{-0.62}^{+1.17}$~Gyr is older than 10 Gyr of
\citet{brown04} by main-sequence photometry. In Figure 4, we plot
the intrinsic SEDs of S312, the integrated SEDs of the
best-fitting model and the spectra of the best-fitting model
superimposed the 3-sigma spectra based on the 2000 Padova
evolutionary tracks. This figure also showed that, the 3-sigma
spectra deviate from best-fitting ones much. As things stand now,
without having other clusters to compare, there is a problem that
an SED-fitting based on the older isochrones (the 1994 Padova)
constrains the age of S312 in agreement with the main-sequence
turnoff fit, while the newer isochrones (the 2000 Padova) give an
older age constraint of S312. These are therefore issues that one
needs to keep in mind in the context of more recent input physical
parameters of the 2000 Padova evolutionary tracks as indicated by
\citet{bru03}. In fact, \citet{bru03} have illustrated the
influence of the stellar evolution prescription on the predicted
photometric evolution of an SSP for solar metallicity as an
example in their Figure 2, which showed models computed using the
1994 Padova, the Geneva and the 2000 Padova evolutionary tracks.
From Figure 2 of \citet{bru03}, we can see that, at late ages, the
$V-K$ color is significantly bluer in the 2000 Padova model than
in the 1994 Padova model. The reason for this is that the
giant-branch temperature in the 2000 Padova tracks has not been
tested against observational calibrations \citep[see][and
references therein]{bru03}. So, it is certain that an SED-fitting
based on the 2000 Padova model gives an older age for S312 than
the age we obtained from the 1994 \citet{bru03} models.

\begin{figure*}
\figurenum{4} \epsscale{1.0} \hspace{0cm}{\plotone{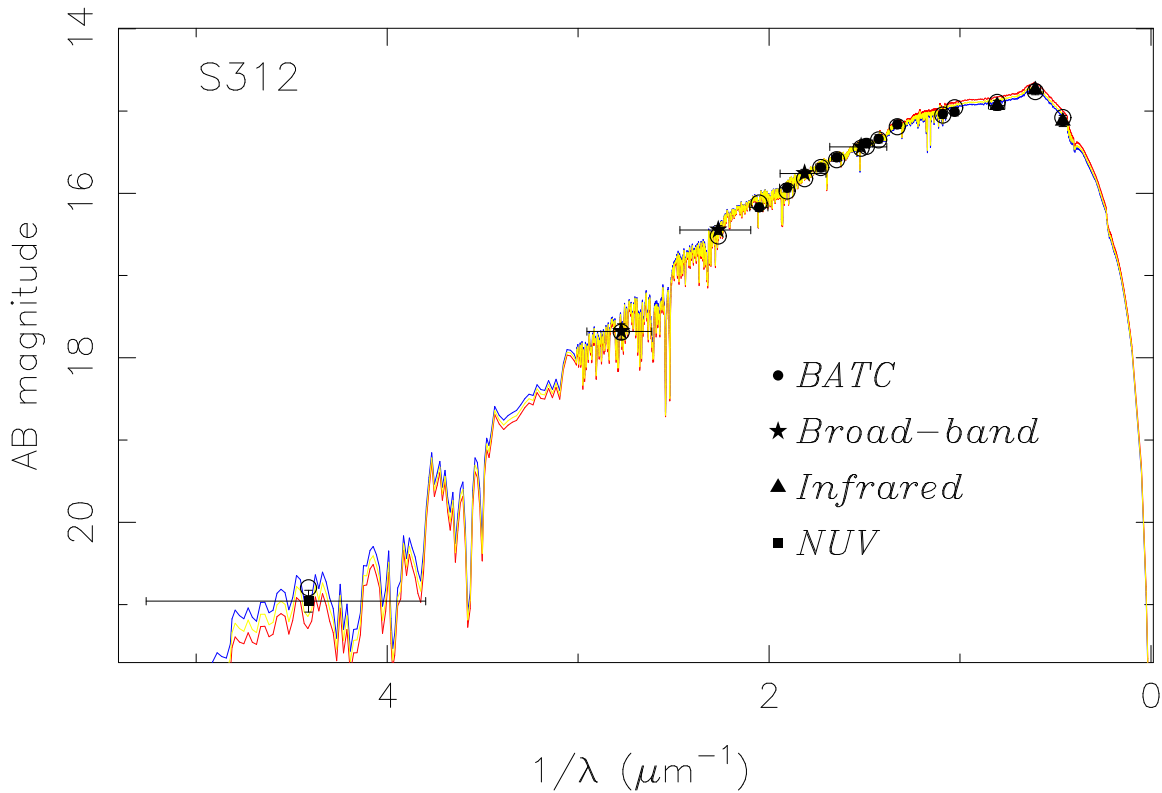}}
\vspace{0.0cm} \caption{Graphs of the best-fit integrated SEDs of
SSP models based on 2000 Padova models plotted on top of the
intrinsic SEDs for S312. The symbols are the same as those in
Figure 3.} \label{fig:fourth}
\end{figure*}

\subsection{Overview of age constraint by photometry in different
bands}

In section 3.4, we constrain the age of S312 based on the
integrated photometric measurements in {\sl GALEX} NUV, broad-band
$UBVR$, 9 intermediate-band filters of BATC and infrared $JHK_s$
of 2MASS. These photometries constitute the accurate SEDs of S312
from $2267-20000$\AA. In order to constrain the age of S312
accurately, we use as many photometric data points covering as
large a wavelength range as possible. In this section, we
investigate which passbands actually constrain the age. The
results show that, except for NUV, a lack of any band does not
result in great problems in constraining the age of S312, the
dispersions in the recovered age are smaller than 0.5 Gyr. The
reason may be that, for S312, whose metallicity and reddening
values were published by other authors, the cluster age is the
sole parameter to be constrained. If we constrain the age and
metallicity of S312 simultaneously, the importance of the $U$ or
$B$ will appear. As \citet{Anders04} pointed out that, the age
deviations from the input values for the combinations without the
$U$ or $B$ bands are caused by an insufficiently accurate
determination of the cluster metallicity, i.e. the $U$ or $B$
plays a major role in determining the metallicity. Missing $U$ or
$B$-band information leads to underestimates of metallicity,
thereby causing ages to be adjusted improperly. The results also
show that, a lack of NUV results in problems in constraining the
age of S312, the dispersion in the recovered age is 2.5 Gyr, which
reflects that the NUV flux point shows promise of being useful
determination of age. A lack of all bands bluer than $V$ or of all
bands redder than $I$ results in great problems in constraining
the age of S312, the dispersions in the recovered age are 4.25 or
7.25 Gyr, respectively.

\begin{figure*}
\figurenum{5} \epsscale{0.45} \hspace{0cm} \rotatebox{-90}
{\plotone{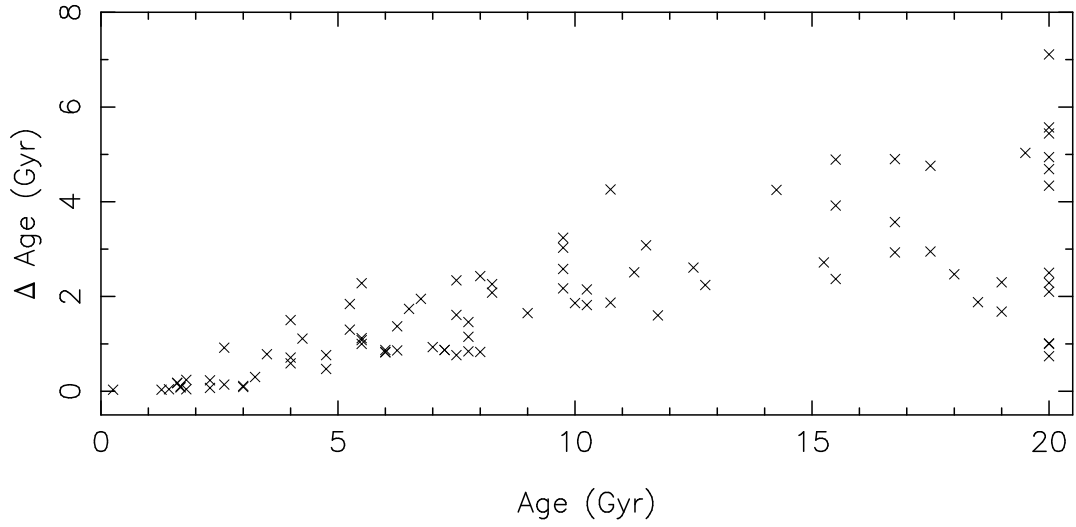}} \vspace{0.0cm} \caption{Age uncertainty
(1$\sigma$ spread) plotted as a function of derived age. The data
are from \citet{fan06}. Notice that ages in excess of 3-4 Gyr are
very uncertain at best.}\label{fig:fifth}
\end{figure*}

\section{Summary and discussion}

In this paper, we constrain the age of the M31 globular cluster
S312 by comparing its multicolor photometry with theoretical
stellar population synthesis models. This is both a check on
previous results, and a check on our methodology of applying age
constraints via our SEDs of stellar systems. Multicolor
photometric data are from {\sl GALEX} NUV, broadband $UBVR$, 9
intermediate-band filters and 2MASS $JHK_s$, which constitute the
SEDs covering $2267-20000$\AA. Our result is in very good
agreement with the previous determination (10 Gyr) of
\citet{brown04} $using$ main-sequence photometry. We also estimate
the mass of S312, which shows that it is a medium-mass globular
cluster in M31. Since no other clusters can be available to
confirm our SEDs-fitting method of constraining ages of globular
clusters, a large sample of globular clusters, whose ages are
derived based on our SEDs-fitting method, may boost confidence in
the dataset and in our SEDs-fitting method. \citet{fan06} obtain
age constraints for 91 M31 globular clusters, based on the same
sources of photometric data (not including {\sl GALEX} NUV), the
same theoretical stellar synthesis models and the same fitting
methods. Figure 5 shows a plot of age errors (1$\sigma$ spread) as
a function of age based on the data of \citet{fan06}. It is clear
that the absolute age errors increase significantly with
increasing age, and derived ages with errors of $<3$ Gyr are
available for ages younger than 9 Gyr.

\citet{fan06} argue for peaks in the age distribution at $\sim$ 3
and 8 Gyr, in addition to the expected complement of Milky
Way-like ``old'' globular clusters. While their results have
considerable scatter, they show a noticeable lack of young,
metal-poor globular clusters, or old and very metal-rich globular
clusters, which might be consistent with an age metallicity
relationship.

At the same time, the results of this paper also indicate that,
despite having such good multicolor photometry, we still cannot
get solid age constraints from SEDs fitting even with NUV to $K_s$
photometry. This has implications for attempting to learn the ages
of distant cluster systems.

\acknowledgments We are indebted to the referee for his/her
thoughtful comments and insightful suggestions that improved this
paper greatly. We thank S.-C. Rey for providing us with his data
on S312 in {\sl GALEX} NUV in advance of publication. S.-C. Rey
also kindly provides us the {\sl GALEX} filter response curves.
This work has been supported by the Chinese National Natural
Science Foundation No. 10473012, 10573020, 10633020, 10673012, and
10603006. This publication makes use of data products from the Two
Micron All Sky Survey, which is a joint project of the University
of Massachusetts and the Infrared Processing and Analysis
Center/California Institute of Technology, funded by the National
Aeronautics and Space Administration and the National Science
Foundation.

\newpage
\begin{table}
\begin{center}
\caption{BATC Photometry of the M31 Globular Cluster S312.}
\label{tab correlation}
\begin{tabular}{ccccc}
\tableline \tableline
Filter & $\lambda$(\AA) & FWHM(\AA) & $N^{a}$ & Magnitude\\
\tableline
   $e$ &4925&390&11& 17.81(0.021)\\
   $f$ &5270&340&12& 17.23(0.016)\\
   $g$ &5795&310&7 & 16.40(0.012)\\
   $h$ &6075&310&5 & 16.14(0.009)\\
   $i$ &6656&480&3 & 15.54(0.007)\\
   $j$ &7057&300&12& 15.25(0.006)\\
   $k$ &7546&330&6 & 14.89(0.006)\\
   $o$ &9182&260&18& 13.95(0.004)\\
   $p$ &9739&270&12& 13.78(0.005)\\
\tableline
\end{tabular}\\
{$^a$ $N$ is the number of images taken by the BATC telescope.}
\end{center}
\end{table}

\begin{table}
\begin{center}
\caption{NUV, broad-band and 2MASS Photometry for
S312.} \label{tab correlation}
\begin{tabular}{lcc}
\tableline \tableline
Filter & Magnitude & Reference \\
\tableline
   NUV   & 21.881(0.123) & \citet{rey06}\\
         &               &              \\
   $U$   & 17.460(0.08)  & \citet{battis87}\\
         &               &              \\
   $B$   & 17.032(0.018) & \citet{rhh94}\\
   $V$   & 16.130(0.011) &\\
   $R$   & 15.560(0.014) &\\
         &               &              \\
   $J$   & 14.195(0.032) & 2MASS        \\
   $H$   & 13.495(0.037) &\\
   $K_s$ & 13.390(0.043) &\\
\tableline
\end{tabular}\\
\end{center}
\end{table}

\end{document}